\documentstyle[graphicx,float]{article}

\begin{document}
\title{Atomic diffraction by light gratings with very short wavelengths}
\date{}
\author{P. Sancho \\ Centro de L\'aseres Pulsados, CLPU \\ E-37008, Salamanca, Spain}
\maketitle
\begin{abstract}
Lasers with wavelengths of the order of the atomic size are becoming
available. We explore the behavior of light-matter interactions in
this emergent field by considering the atomic Kapitza-Dirac effect.
We derive the diffraction patterns, which are in principle
experimentally testable. From a fundamental point of view, our
proposal provides an example of system where the periodicity of the
diffraction grating is comparable to the size of the diffracted
object.
\end{abstract}

\section{Introduction}

Lasers in the X-ray domain, with wavelengths of the order of the
atomic size, have been reported in the literature
\cite{Nat,xfe,All}. We must study the underlying physics in that unexplored
range of ultrashort wavelengths. In particular, we must analyze the
behavior of the light-matter interaction. In
order to carry out the analysis in a simple way, we shall consider a
well-known scheme, the Kapitza-Dirac effect, which can be described
with simple mathematical tools. In the Kapitza-Dirac effect a beam
of atoms or electrons is diffracted or scattered by a standing light
wave \cite{KD, Bat, Pri, Ba2,Ba3}. The effect can be extended to
two-particle systems \cite{Sa1,Sa2}.

First of all, we must derive the form of the light-atom interaction
in this regime. We shall show that the dynamics is ruled by a
lightshift potential in the dipole approximation. Being the light
wavelength comparable to the atomic size, different parts of the
atom feel different values of the field and the dipole approximation
provides an incomplete description of the problem. We must consider
higher multipole terms, in particular the quadrupole one. However,
because of the high frequency of the light field we must average the
interaction on time. The average of the higher permanent multipole
terms is zero, reducing the total interaction to the usual
lightshift potential in the dipole approximation. We shall also consider
the quadrupole induced by the electric field. The evaluation of
its actual contribution to the problem is difficult because of the
lack of reliable experimental data on quadrupole or higher order
polarizabilities. Nevertheless we shall show that if these effects
were of the same order of the lightshift potential they would not
modify the fundamental results of our paper.

In the standard atomic Kapitza-Dirac arrangement the wavelength of
the optical grating is close to an atomic transition, enhancing the
strength of the interaction. The very short wavelengths we consider
in this paper are fully detuned from the atomic transitions, making
much more weak the light-atom interaction, which is only related to
the atomic polarizability. However, considering a laser intensity of
the order of $10^{14}W/m^2$ (the values used in the observation of
the effect with electrons \cite{Ba2}) the strength of the light-atom
interaction is comparable to that in the on resonance case and we
can, in principle, observe the diffraction effects. We shall
evaluate the diffraction patterns, obtaining the same form of the
standard atomic Kapitza-Dirac arrangement (containing only even
order peaks).

The high frequency values of the light can give rise, depending on
the duration of the interaction, to the presence of large ionization
rates which would make more difficult the observation of the
diffraction effects. In these cases one must introduce some
procedure to remove the ionized atoms from the experiment.

In addition to provide an example of light-matter interaction in the
scale of very short wavelengths, our proposal is also interesting
from a fundamental point of view. There has been an increasing
interest in the study of quantum diffraction with large size
objects, for instance, with $C_{60}$ \cite{Ze1} and $C_{70}$
\cite{Ze2} molecules in solid nanostructures or structures made of
light \cite{Ze3}, and $Na_2$ molecules whose de Broglie wavelength
is smaller than their size \cite{Pr1}. However, there is yet a
fundamental question that has not been addressed, the limiting size
(relative to the spacing of the diffraction grating) of a quantum
object to observe diffraction. Our proposal can be the basis to
study quantum diffraction in the extreme regime where the
periodicity of the diffraction optical grating is similar to the
size of the diffracted system.

The plan of the paper is as follows. In Sect. 2 we present the
arrangement and evaluate the lightshift potential relevant for the
problem. In order to give a compact presentation of the main ideas
involved in the evaluation, some more technical aspects of the
treatment are discussed in Appendix 1. In Sect. 3 we derive the
diffraction patterns and consider how to eliminate the ionized
atoms. The values of the parameters involved in a realistic
experimental implementation of the arrangement are estimated in
Sect. 4. In the Discussion, we consider the potential impact of our
proposal for other physical problems, in particular, the exploration
of the limits of quantum diffraction with large size objects.
Finally, the possible effects associated with a large quadrupole
polarizability are presented in Appendix 2.

\section{The arrangement}

As usual in the Kapitza-Dirac effect our arrangement consists of a
standing light wave generated by a laser. A beam of atoms interacts
with that diffraction grating. Behind the grating we place
detectors. From the data collected at the detectors we can infer the
diffraction patterns (see Fig. 1).

Next, we consider the mathematical description of the interaction.
First of all, we note that we shall deal with high laser intensities.
Then we can resort to the semiclassical approximation, where the
electromagnetic fields can be treated classically. In Appendix 1
we estimate the intensity values for which we can safely use this
approximation in the framework of the Kapitza-Dirac effect.
\begin{figure}[H]
\center
\includegraphics[width=5cm,height=3cm]{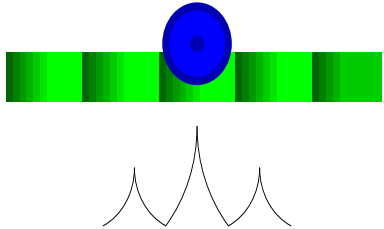}
\caption{The wavelength of the laser (green) is similar to the size
of the atom (blue). After the interaction we observe a diffraction
pattern.}
\end{figure}
The adiabatic condition plays an important role in this type of
problem. When it is fulfilled, the time dependence of the effective
optical potential experienced by the atoms is slow compared to the
internal evolution. The center of mass (CM) dynamics can be
decoupled from the internal one, and it can be described by an
optical potential, generating a phase shift as a function of
position \cite{Ada}. As we shall see later (Sect. 3), the adiabatic
condition is by far fulfilled in our case.

Once guaranteed adiabatic evolution, we can concentrate on the CM
dynamics. The phase shift of the CM wavefunction is given by the
dipole-type interaction between the electric field ${\bf E}$ and the
induced atomic dipole ${\bf d}=\alpha {\bf E}$, with $\alpha $ the
polarizability. We describe this interaction by the lightshift
potential \cite{Bat,Ze3}
\begin{equation}
U_{LS}({\bf R},t)=-\frac{1}{2} \alpha {\bf E}^2({\bf R},t)
\end{equation}
where ${\bf R}$ denotes the CM coordinate of the atom. Being the
CM timescale very different from the laser period, we must average
over that period. If we take for the electric field of the standing
wave the form
\begin{equation}
{\bf E}({\bf R},t)={\bf E}_0 \cos ({\bf k}_L \cdot {\bf R}) \cos (\omega _L t)
\end{equation}
with ${\bf k}_L$ and $\omega _L$ the wavevector and frequency of the
laser, the averaged lightshift potential takes the form
\begin{equation}
U({\bf R})= \frac{1}{T_L}\int _0^{T_L} U_{LS}dt =-\frac{1}{4} \alpha {\bf E}_0^2 \cos ^2({\bf k}_L \cdot {\bf R})
\end{equation}
with $T_L=2\pi /\omega _L$.

In the framework of the dipole approximation there is yet another
interaction channel (see Appendix 1). In addition to the
transitions between bound states that induce the atomic dipole,
there can be also transitions to the continuum. As a matter of fact,
due to the high energy of the involved photons we expect that
absorption will lead in most cases to ionization, and the ionization
rate in our problem can be large. This is not the case for the
values of the parameters we shall propose to observe the effect (see
Sect. 3). However, for other values of the parameters the ionization
rate could be large. In these cases, as we are only interested into
the diffraction properties of the non-ionized atoms, we must
consider an scheme to eliminate the ionized atoms. In the next
section we present such an scheme. Thus, due to the negligible value
of the ionization rate in some cases and to the possibility of
eliminate the ions in the rest of cases, we do not need to consider
this interaction channel.

Up to now we have restricted our considerations to the leading
dipole approximation. However, as remarked before, because the
optical wavelength is comparable to the atomic size, the dipole
approximation (which assumes no relevant variations of the field in
distances of the order of the atomic size) does not provide a
complete description of the problem. There are in the literature
several examples of how to go beyond this approximation in the
context of X-ray theory. For instance, quadrupole terms have been
used to study X-ray spectroscopy \cite{Ber}, numerical non-dipole
simulations of ionization by X-ray lasers have been presented in
\cite{Zho}, and the Bloch equations without the dipole approximation
have been derived in \cite{Zha}. In this paper we follow the
approach of considering higher multipole terms. Let us consider the
next term in the perturbative series, the permanent quadrupole term,
which has the form (see Appendix 1)
\begin{equation}
U_Q({\bf R},t)=\frac{1}{2} Q_{ij} \frac{\partial E_i}{\partial R_j}
\end{equation}
Using the form of the electric field we have that $U_Q \sim \sin
({\bf k}_L \cdot {\bf R}) \cos (\omega _L t)$. Now, when performing
the time average we have that the quadrupole interaction goes to
zero. Thus, although there is a quadrupole interaction associated
with the field variations along the atom size, its net effect
vanishes. As discussed in Appendix 1 a similar conclusion holds for
any higher order term. Thus, all the averaged multipole permanent
contributions vanish and the light-atom interaction can be described
via the dipole approximation. In Appendix 2 we shall discuss how
induced quadrupole multipoles could be present in the problem.

In conclusion, the analysis of this section shows that the light-atom interaction can be expressed in the form
\begin{equation}
U({\bf R})=U_0 \cos ^2 ({\bf k_L} \cdot {\bf R})
\end{equation}
with $U_0 = -\alpha {\bf E}_0^2/4$. This is the usual lightshift
potential used in the standard on resonance atomic Kapitza-Dirac effect.

\section{Predicted diffraction pattern}

We have shown in the previous section the existence of two
interaction channels in our arrangement. On the one hand, the large
energy of the photons in the range of frequencies considered can
lead, depending on the duration of the interaction, to a high
ionization rate. On the other hand, the atoms that are not ionized
generate a diffraction pattern. We are only interested into the
observation of these  patterns. Then in the cases where the
ionization is large we must remove the ionized atoms (and the
electrons) in order to postselect the neutral ones.

One can easily design methods able to extract the ionized atoms and
electrons from the experiment. For instance, an electric field
perpendicular to the plane where the experiment takes place
(perpendicular to the longitudinal and transversal directions) would
deviate most electrons and ions from the detectors, whereas the
evolution of the neutral atoms would be almost unaffected.

After the removal of ions and electrons we can focus on the dynamics
of the neutral atoms interacting with the lightshift potential. The
state of the CM at time $t$ of the atoms with initial wavevector
$k_0$ (at time $t=0$) can be expressed as $e^{iU(X)t /\hbar}
e^{ik_0X}$ in the Raman-Nath approximation. In this approximation we
assume that the momentum of the atoms is large compared to that of
the photons. Then the kinetic energy remains approximately constant
and may be neglected. As it is well-known, this approximation is
valid in the diffraction regime \cite{Bat}. The coordinate $X$ is
that of the CM in the direction parallel to the granting. As usual
in the Kapitza-Dirac effect we only consider the one-dimensional
problem \cite{Bat}.

Using the relation $\exp (i\xi \cos \varphi)=\sum _n i^n J_n(\xi
)\exp (in\varphi )$, with $J_n$ the n-th order Bessel function, we
have
\begin{equation}
e^{iU(X)\tau /\hbar} e^{ik_0X}= e^{iU_0\tau /2\hbar} \sum _{n=-\infty}^{\infty } i^n J_n \left( \frac{U_0 \tau }{2\hbar } \right) e^{i(2nk_L + k_o)X}
\end{equation}
with $\tau $ the interaction time.

This pattern shows the standard form in Kapitza-Dirac diffraction
(that with on resonance light-atom interaction). Only even
diffraction orders are present. The intensity of the peaks is given
by $ J_n ( U_0 \tau /2\hbar )^2$.

\section{Experimental parameters}

We analyze in this section the values of the parameters of the
problem that would lead to observable effects in an experimental
realization of the above arrangement. We shall use a wavelength
$\lambda _L \approx 5 \times 10^{-10}m$. This wavelength differs in
several orders of magnitude of those associated with atomic
transitions, giving rise to a large detuning, which guarantees the
adiabatic evolution in the problem.

First of all, in order to be in the diffraction regime the
coefficient $U /\epsilon $, with $\epsilon =\hbar ^2k_L^2/2m$ the
recoil shift of the atom by absorption of a photon, must be larger
than unity \cite{Bat}. Taking the mass of the atom as ten to twenty times the
proton mass, we have $\epsilon \approx 10^{-4} eV$, and we
must take $U \approx 10^{-3}eV$.

From this value of the potential we can deduce the intensity of the
laser beam. We use the approximate relation $U \approx \alpha E_0^2$
and the definition of intensity $I=cE_0^2/8\pi $. The atomic
polarizability values are in the range of $10^{-29}m^3$ to
$10^{-31}m^3$ \cite{Sch}. Taking an atom in the high part of the
range, $\alpha \approx 10^{-29}m^3$, we would need a laser intensity
of the order of $10^{14}W/m^2$. This can seem a very high
value when compared with the intensities in the standard atomic
Kapitza-Dirac effect ($I \approx 10^7 W/m^2$). However,  a value of
$5 \times 10^{14}W/m^2$ has been used to demonstrate the effect with
electrons \cite{Ba2}. The proposed value of $\lambda _L =5 \times
10^{-10}m$ has already been reached with beam intensities as high as
$10^{17}W/m^2$ \cite{Nat,xfe}. Note also that the proposed value, although
very high, is yet a long way from the threshold of non-linear
polarizabilities, around $I=10^{18}W/m^2$.

The condition of high visibility of the interference pattern, $U\tau
/\hbar \approx 1$ \cite{Bat}, implies a time of interaction of
around $10^{-12}s$. As the duration of the pulses in \cite{Nat,xfe}
is between $10^{-13}$ and $10^{-14}s$, a possibility to generate the
standing wave is using counter-propagating pulses slightly enlarged
on time (for instance, via dispersion of the ultrashort pulses).
Another possibility, which does not modify the pulse duration, is to
increase the velocity of the atoms. The smallest radios of the
focused spots for this range of wavelengths are around $1\mu m$
\cite{Nat}. Then $\tau \approx 10^{-12}s$ requires atom velocities
close to $10^6 ms^{-1}$, that is, an increase of three orders of
magnitude with respect to the usual values. These velocities could
be reached accelerating ions, which later would interact with free
electrons generating neutral atoms (the remaining ions should be
extracted from the beam by interaction with an electric field). In
this second scheme one cannot use counter-propagating laser beams,
which are shorter than $\tau$, and should generate the standing wave
with mirrors or other techniques.

We must also determine the fraction of atoms that are ionized during
the interaction. If a photon is absorbed the probability of
ionization is large in our arrangement. For $\lambda _L$ the energy
of one photon is $E_{ph}=\hbar \omega _L \approx 3 \times 10^2 eV$.
This energy is much larger than the ionization energy, giving rise
to a large probability of ionization in the case of photon
absorption. We estimate the number of non-ionized atoms, $N$, in the
usual way: $dN =-\Gamma N dt$ with $\Gamma $ the ionization rate,
which is assumed to be time-independent. This gives $N(\tau )=N_0
\exp (-\Gamma \tau)$ with $N_0$ the initial number of atoms in the
beam. On the other hand, for the range of intensities in our problem
the ionization rate can be expressed as $\Gamma =\sigma I/\hbar
\omega _L$ with $\sigma$ the photoionization cross section, which
depends on $\omega _L$. Values of $\sigma $ can be found in several
database. For instance for the Na atom and $\hbar \omega _L = 100eV$
we have $\sigma =5 \times 10^{-22} m^2$, which gives $\Gamma \tau
\approx 10^{-3}$. Due to the very short duration of the interaction
the fraction of ionized atoms is negligible. For the values of the
parameters here proposed it is not necessary to include the
procedure to remove the ionized atoms.

\section{Discussion}

Some coherent light sources are reaching the scale where the optical
wavelengths are comparable to the atomic sizes. Our work is one of
the first studies considering possible physical effects present in
this unexplored regime. Contrarily to a naive intuition, the
effective interaction does not depend on multipole terms beyond the
dipole one, and the evolution can be described via the usual
lightshift potential. We have shown the existence of diffraction
effects similar to those present when the light is on resonance with
atomic transitions. The main obstacle to carry out the experiment is
the  short duration of the laser pulse. For other possible
experiments with much longer durations (for instance, Bragg's
scattering with very short wavelengths) the ionization rates can be
high and a procedure to eliminate the ionized atoms should be added
to the arrangement.

In addition to the interest of our proposal in atom optics, it could
also be relevant to other fields. We shall focus on three of them.
The first one concerns to fundamental physics, in particular, to the
understanding of the wave properties of quantum systems. Our scheme
provides the first example where the periodicity of the classical
diffraction grating is similar to the size of the quantum diffracted
object. This is an extreme scenario for quantum diffraction that
extends the research presented in \cite{Ze1,Ze2,Ze3,Pr1}. Our
analysis shows that in the limit of objects of the size of the grating
periodicity, diffraction survives.

Also, from a fundamental point of view, our proposal serves to test the
theory of light-matter interactions in the scale of very short
wavelengths. This knowledge would be necessary to study, for
instance, the dispersion of this type of light by atoms. It should
be also in the basis of techniques aimed to provide images of the
spatial structure of atoms, a process in principle possible with
light of the same wavelength of the size of the illuminated object.

Finally, from a more practical point of view, the diffraction of
atoms can be used to measure the polarizability, by
fitting the experimental data to the detection distributions.
This would be an alternative method of measurement of this atomic property.

{\bf Acknowledgments} I am grateful to Luis Plaja for discussions on
the problem. I acknowledge support from Spanish Ministerio de
Ciencia e Innovaci\'on through the research project FIS2009-09522.

{\bf Appendix 1}

In this Appendix we present some technical points that complement
the main developments in the paper.

{\it Semiclassical approximation.} We derive the values of the laser
intensity that guarantee the use of the semiclassical approximation.
The usual criterion for its validity is to have a large number of
photons in the interaction region. In standard Kapitza-Dirac
diffraction, where the approximation works well, the number of
photons per unit volume is $I/c\hbar \omega _L \approx 10^{18}
photons/m^3$, where we have used $\lambda _L =500 nm$ and $I=10^7
W/m^2$. The typical volume of the interaction region is $10^{-12}
m^3$ ($10 \mu m$ (aperture collimating the beam of atoms) $\times$
$1 mm$ (height of the laser beam) $\times$ $100 \mu m$ (width of the
laser beam)). The number of photons in this characteristic volume is
around $10^{6}$. For short wavelengths, the energy of each photon is
around $10^2 eV$ and, in order to have a similar number of photons
(assuming a similar characteristic volume), the intensity must be
above $10^{11} W/m^2$. Then in our proposal we can safely use the
semiclassical approximation.

{\it Multipole expansion.} For very short wavelengths different
parts of the atom can feel different electric fields. We must go
beyond the dipole approximation and consider higher multipole terms.
The light-atom interaction potential can be expressed in the
multipole form \cite{Lou}:
\begin{equation}
{\bf D}\cdot {\bf E}({\bf R},t) + \frac{1}{2} \sum _{i,j} Q_{ij} \left( \frac{\partial E_i}{\partial r_j} \right) ({\bf R},t) + \cdots
\label{eq:mul}
\end{equation}
where ${\bf D}$ and $Q_{ij}$ represent the dipole and $ij$-component
of the quadrupole momenta of the atom. They can be rewritten using
the second quantization of the atomic variables \cite{Lou}. If we
denote by $|h>$ and $|l>$ the eigenstates of the atom we have
$\hat{\bf D}=\sum _{l,h}{\bf d}_{lh}|l><h|$ with ${\bf D}_{lh}=
\sum _{e_n} <l|e_n {\bf r}_{e_n}|h>$, and $\hat{Q}_{ij}=\sum _{l,h} Q_{ij})_{lh}|l><h|$ with $(Q_{ij})_{lh}= \sum _{e_n} <l|e_n ( 3(r_{e_n})_i(r_{e_n})_j - \delta _{ij}{\bf r}_{e_n}^2)|h>$. The sum, $\sum _{e_n}$, is over all the electrons of the atom. The diagonal elements (expectation values) ${\bf D}_{ll}$ are null because the integrand has odd parity. The non-diagonal (transition) elements contribute to the atomic polarizability. In the static and
spherically symmetric case the polarizability has the form
\begin{equation}
\alpha =\frac{2}{3} \sum _{l \neq g} \frac{|<g|\sum _{e_n}{\bf r}_{e_n} |l>|^2}{{\cal E}_l-{\cal E}_g}
\end{equation}
with $g$ denoting the ground state, and ${\cal E}_l$ he energy of the state
\cite{Sch}.

The dipole term also describes the ionization of the atom. It is an
alternative channel associated with ground-continuum transitions,
with matrix elements in the form $<c|e{\bf r}|g>$, denoting $|c>$
states of the continuum.

The next term in Eq. (\ref{eq:mul}) represents the permanent
quadrupole effects. In general, the values of the permanent
quadrupole momenta can be expressed as $Q_{ij} \sim er_0^2$, with
$r_0$ Bohr's radius and a proportionality coefficient of the order
of unity \cite{Ita}. As signaled before, in order to be in the
diffraction regime the potential must be of the order of $U \approx
10^{-3}eV$. Then the quadrupole term must be similar, $U_Q \approx
10^{-3}eV$. Taking for the field the standard form we can express
the potential as $U_Q \approx er_0^2k_LE_0 \approx er_0E_0$. Using
this expression and the relation between $I$ and $E_0$ we have that
the intensity of the laser must be around $10^{8} W/m^2$ in order to
the quadrupolar effects to be relevant in the problem. This value is
smaller than the intensity required to have a non-negligible atomic
dipole induced by the polarizability. In Appendix 2 we discuss the
role of induced quadrupole terms in our problem.

Note that if the n-th order permanent multipole coefficient follows a relation
of the type $Q^n \sim er_0^n$ with a proportionality coefficient of
the order of unity, we have $U_{Q^n} \approx er_0^nk_L^{n-1}E_0
\approx er_0E_0$ and that multipole term also has to be taken into
account. However, as in the quadrupole case, its temporal dependence
remains in the form $\cos(\omega _L t)$, and after averaging it can
be neglected.

Finally, we recall the fact that the electric quadrupole and the
magnetic dipole energies have similar magnitudes. Thus, if the
electric quadrupole term is relevant for the problem, the magnetic
dipole one must also be taken into account. The magnetic dipole term
reads as $e{\bf D}_M \cdot {\bf B}({\bf R})$ with ${\bf B}({\bf R})$
the magnetic field at the atomic CM position and ${\bf D}_M$ the
magnetic dipole moment, which can be expressed as ${\bf
D}_M=-(1/2m)\sum _i {\bf l}_i$ with ${\bf l}_i$ the orbital angular
momentum of the i-th electron in the atom. From these expressions
and the form of the electromagnetic potential it is clear that the
temporal dependence of this term is also proportional to $\cos
(\omega _L t)$, and it will vanish after time averaging just as in
the electric quadrupole case.

{\bf Appendix 2} In this Appendix we analyze the atomic quadrupole
induced by the electric field or its gradient. When the
polarizability effects are taken into account the quadrupole momenta
can be expressed as
\begin{equation}
Q_{ij}=Q_{ij}^0 + \sum _k A_{kij}E_k + \sum _{kl} C_{ijkl} \frac{\partial E_k}{\partial x_l}
\end{equation}
where $ Q_{ij}^0$ are the permanent momenta, $A_{kij}$ is the
dipole-quadrupole polarizability and $C_{ijkl}$ the
quadrupole-quadrupole one. The first term in the r. h. s. of the
equation represents the quadrupole momenta in absence of external
electric fields, the second the momenta induced by an electric field
and the third these induced by the gradient of the field.

The order of magnitude of the coefficients is $A \approx
e^2r_0^3E_h^{-1}$ and $C \approx e^2r_0^4E_h^{-1}$ with $E_h = 4
\times 10^{-18} J$. Introducing numerical values we obtain for the
potentials associated with these two terms ($U_A
\approx AE_0k_LE_0$ and $U_C \approx C(E_0k_L)^2$): $U_A \approx U_C \approx (er_0E_0)^2 E_h^{-1} \approx 10^{-4} - 10^{-5} eV $, where we have used $r_0 k_L \approx 1$. We expect the potentials to be ten to one hundred times smaller than the lightshift potential. However, to extract precise conclusions of the importance of the induced terms we should know the actual values of the coefficients. Unfortunately, very few experimental data are available, and in many cases there are large uncertainties on their theoretical estimation (see, for instance, \cite{Por} for some recent work in this subject. Note that with the values suggested for Mg in this reference the induced potential would be comparable to the lightshift one).

In our particular problem, when the quadrupole polarizability is
taken into account the full potential can be written as
\begin{eqnarray}
U(X)=U_0 \cos ^2(k_LX) + U_A \cos ^3 (k_LX)\sin (k_L X) \nonumber
\\ + U_C \cos ^2 (k_LX)\sin ^2 (k_L X)
\end{eqnarray}
Note that the temporal dependence in all the terms of the r. h. s.
is $\cos ^2 (\omega _L t)$. All the terms remain after the time averaging.
The coefficient resulting from the integration is included in $U_0,
U_A, U_C$. Using simple trigonometric relations we obtain
\begin{eqnarray}
U(X)=\frac{U_0}{2}+\frac{U_C}{8} + \frac{U_0}{2} \cos (2k_LX)+
\frac{U_A}{4} \sin (2k_LX) \nonumber \\ + \frac{U_A}{8} \sin
(4k_LX)- \frac{U_C}{8} \cos (4k_LX)
\end{eqnarray}
The final detection pattern is
\begin{eqnarray}
e^{iU(X)\tau /\hbar} e^{ik_0X}= e^{i\left( \frac{U_0}{2} +
\frac{U_C}{8} \right) \tau /\hbar} \sum _{n,m,l,r=-\infty}^{\infty }
i^{n+r} J_n \left( \frac{U_0 \tau }{2\hbar } \right) \times
\nonumber \\ J_m \left( \frac{U_A \tau }{4\hbar } \right) J_l \left(
\frac{U_A \tau }{8\hbar } \right) J_r \left( -\frac{U_C \tau
}{8\hbar } \right) e^{i([2n+2m+4l+4r]k_L + k_o)X}
\end{eqnarray}
where we have used the relation $\exp (i\xi \sin \varphi)=\sum _n
J_n(\xi )\exp (in\varphi )$. The diffraction pattern shows the same
analytical form found for the lightshift potential, the peaks
correspond to even multiples of $k_L$. The difference lies in the
different intensities of the peaks, which in addition to $U_0$ now
also depend on $U_A$ and $U_C$. In conclusion, in presence of
quadrupole polarizability effects, we would obtain a diffraction
pattern similar to that associated with dipole polarizability.
Our main result remains valid, there are diffraction patterns in the
regime of very short wavelengths. In addition, the diffraction
pattern could be used to determine the $U_A$ and $U_C$ values and,
in consequence, the quadrupole polarizabilities, which are very difficult to
measure by other methods \cite{Por}.

\end{document}